\documentclass[11pt]{article}
\usepackage{axodraw,amsmath,epsfig}
\usepackage{epsfig} 
\usepackage{amssymb}
\usepackage{bbm}
\newsavebox{\PSLASH}
\sbox{\PSLASH}{$P$\hspace{-2mm}/}

\begin{document}
\setcounter{footnote}{0}
\renewcommand{\thefootnote}{\alph{footnote}}
\renewcommand{\theequation}{\thesection.\arabic{equation}}
\newcounter{saveeqn}
\newcommand{\add}{\addtocounter{equation}{1}}
\newcommand{\alpheqn}{\setcounter{saveeqn}{\value{equation}}%
\setcounter{equation}{0}%
\renewcommand{\theequation}{\mbox{\thesection.\arabic{saveeqn}{\alph{equation}}}}}
\newcommand{\reseteqn}{\setcounter{equation}{\value{saveeqn}}%
\renewcommand{\theequation}{\thesection.\arabic{equation}}}

\newsavebox{\DELVECRIGHT}
\sbox{\DELVECRIGHT}{$\stackrel{\rightarrow}{\partial}$}
\newcommand{\PARVECR}{\usebox{\DELVECRIGHT}}
\thispagestyle{empty}
\begin{flushright}
\par
\end{flushright}
\vspace{1.5cm}
\begin{center}
{\Large\bf{Secondary particles spectra in the decay of a polarized top quark with anomalous $tWb$ couplings}}\\
\vspace{1cm}
\bf Mojtaba Mohammadi Najafabadi\footnote{{Electronic
 address:
mojtaba.mohammadi.najafabadi@cern.ch}}  \\ 
\hspace{0.2cm} \\

\vspace{0.5cm}
{\sl  Department of Physics, Sharif University of Technology}\\
{\sl P.O. Box 11365-9161, Tehran-Iran}\\
and\\
{\sl Institute for Studies in Theoretical Physics and Mathematics (IPM)}\\
{\sl{School of Physics, P.O. Box 19395-5531, Tehran-Iran}}\\

\end{center}
\vspace{0cm}
\begin{center}
{\bf {Abstract}}
\end{center}
\begin{quote}
Analytic expression for energy and angular dependence of a secondary
charged lepton in the decay of a polarized top quark with 
anomalous $tWb$ couplings in the presence of all anomalous couplings are derived. 
The angular distribution of the b-quark is derived as well. 
It is presented that the charged lepton spin correlation coefficient is not very 
sensitive to the anomalous couplings. However, the b-quark spin correlation coefficient is 
sensitive to anomalous couplings and could be used as a powerful tool for searching of 
 non-SM coupling.
\end{quote}
\hspace{0.8cm}
\par\noindent

\newpage
\setcounter{page}{1}
\section{Introduction}
Studying top quark is particularly interesting for several practical reasons. Indeed due to 
its large mass the electroweak gauge symmetry is broken maximally by top quark and 
it may be the first place in which the non-SM effects could be appeared.
Such a large mass leads to a very short life time for the top 
quark ($\tau_{top} \sim 4\times 10^{-25}$ s).
This number is roughly one order of magnitude smaller than the typical QCD hadronization
time scale ($3\times 10^{-24}$ s). The top quark therefore decay before it can hadronize
unlike the other quarks.
As a result, the spin information of the top quark is transferred to its decay products.
In the case of having an appropriate choice of spin quantization axis 
the top quark spin can be utilized as a strong tool for investigation of new physics in the top
quark interactions.
But the rate of ``polarized'' top quark production at LHC and Tevatron is very small. 
For this reason, the existing 
correlations between top spin and anti-top spin in the $t\bar{t}$ production and the 
correlation between top spin 
and the direction of the light quark momentum in the single top t-channel 
production (which is the largest source of single top at the LHC)
can be used to reconstruct the spin information of the top quark \cite{Beneke}.
There have been some studies in probing of the anomalous $tWb$ couplings in the past, for example
in \cite{Boos1},\cite{Boos2},\cite{nakano}.
The effect of anomalous $tWb$ couplings on angular distribution of secondary particles has been
studied in \cite{Hioki1},\cite{Hioki2},\cite{Hioki3}.
However, in those studies all quadratic contributions of the anomalous couplings 
have been neglected and the narrow width approximation for the  W-boson has been used.
 
In this paper, the effects of the anomalous $tWb$ couplings on the spectra 
of a secondary charged lepton
and b-quark in the decay of a polarized top quark are shown by obtaining the analytic expression.
The dependency of the top width on the anomalous couplings is presented too.
In particular, the dependency of spin correlation coefficients (which are observables in top events
at hadron colliders) on the anomalous couplings
for the charged lepton and b-quark are derived.
This calculation has been performed in the presence of the all
anomalous couplings $(f_{R1},f_{L2},f_{R2})$ and
quadratic contributions of anomalous couplings are kept.
Regarding to the bounds obtained in \cite{Boos1} and \cite{nakano}
the quadratic terms seem to be really important. For instance, the $95\%$ C.L. estimated
bounds on a ratio of anomalous couplings using $t\bar{t}$ events are \cite{nakano}:
\begin{eqnarray}
&&-0.93<\frac{f_{R2}}{f_{L1}}<0.57  ~~ (\text{for 100 reconstructed events at Tevatron})\nonumber \\ 
&&-0.81<\frac{f_{R2}}{f_{L1}}<-0.70 ~,~  -0.12<\frac{f_{R2}}{f_{L1}}<0.14 ~~ 
(\text{for 1K reconstructed events at Tevatron})\nonumber \\ 
&&-0.74<\frac{f_{R2}}{f_{L1}}<-0.72 ~,~  -0.01<\frac{f_{R2}}{f_{L1}}<0.01 
~~ (\text{for 100K reconstructed events at the LHC}) \nonumber
\end{eqnarray}
In obtaining these constraints the detector effects have  been simulated taking into account 
the geometry and energy resolutions of CDF and ATLAS detectors for the Tevatron and the LHC, respectively.
The above bounds are still optimistic values because only the statistical uncertainties have been taken into account 
and the systematic uncertainties neglected.

The $95\%$ C.L. estimated
bounds on the anomalous couplings using single top events with taking into account $10\%$ 
of systematic uncertainties are \cite{Boos1}:
\begin{eqnarray}
&&-0.24<f_{R2}<0.25  ~,~-0.18<f_{L2}<0.55 ~~~\text{Tevatron}\nonumber \\ 
&&-0.17<f_{R2}<0.18  ~,~-0.094<f_{L2}<0.34 ~~~\text{LHC}\nonumber 
\end{eqnarray}

From measurements of the FCNC processes of rare B decays, 
$b\rightarrow s\gamma$,
the CP-violation is severely constrained \cite{CP}, which means that the anomalous couplings
can be taken as real values.
The ``right-handed'' top quark coupling is strongly constrained \cite{belle}. However, it is kept 
in the calculations in this paper to have the most general expression and show its effect.

This paper is organized as follows:
In section 2 the effective Lagrangian of $tWb$ anomalous couplings is introduced.
Section 3 is dedicated to present some results
of two-body and three-body decay of the top quark with anomalous Lagrangian
introduced in section 2 and finally section 4
concludes the paper.

\section{The effective $tWb$ Lagrangian}
The Lagrangian of new physical phenomena which may occur at the energy scale of $\Lambda$
 can be expanded in terms of ($1/\Lambda$) as:

\begin{eqnarray}\label{effLag}
{\mathcal L}_{eff} = {\mathcal L}_{SM} + \frac{1}{\Lambda^{2}}\sum_{i}C_{i}O_{i} + {\mathcal O}(\frac{1}{\Lambda^{4}}) , 
\end{eqnarray}
where ${\mathcal L}_{SM}$ is the SM Lagrangian and $O_{i}$ are the SM-gauge-invariant dimension-six operators.
$C_{i}$ represent the coupling strengths of $O_{i}$ \cite{Lag},\cite{Peccei}.
Keeping the $CP$ conserving terms in dimension six in the unitary gauge leads to the following Lagrangian \cite{Peccei}:
\begin{eqnarray}\label{intLag}
L = \frac{g}{\sqrt{2}}[W^{-}_{\mu}\bar{b}\gamma^{\mu}(f_{L1}P_{L} + f_{R1}P_{R})t - \frac{1}{2M_{W}}W_{\mu\nu}
\bar{b}\sigma^{\mu\nu}(f_{L2}P_{L} + f_{R2}P_{R})t] + h.c. ,
\end{eqnarray}
where $W_{\mu\nu} = D_{\mu}W_{\nu} - D_{\nu}W_{\mu}$, $D_{\mu} = \partial_{\mu} - ieA_{\mu}$,
$P_{R,L} = \frac{1 \pm \gamma_{5}}{2}$ and $\sigma_{\mu\nu} = \frac{i}{2}(\gamma_{\mu}\gamma_{\nu} -
\gamma_{\nu}\gamma_{\mu})$.

The corresponding feynman rule is:

\vskip0.2cm
\hspace{0.0cm}
\begin{eqnarray}\label{Gamma}
\SetScale{1}
    \begin{picture}(50,20)(0,0)
    \ArrowLine(-30,0)(0,0)
    \Vertex(0,0){2}
    \Photon(0,0)(30,-30){2}{4}
    \ArrowLine(0,0)(30,30)
    \Text(20,35)[]{$b$}
    \Text(-25,-10)[]{$t$}
    \Text(14,-30)[]{$W^{+}_{\mu}$}
    \end{picture}
\hspace{0.0cm}
\lefteqn{\Gamma^{\mu} = -\frac{g}{\sqrt{2}}[\gamma^{\mu}(f_{L1}P_{L} + f_{R1}P_{R}) }\nonumber\\&&\hspace{0.2cm}
- i\sigma^{\mu\nu}(p_{t} - p_{b})_{\nu}
(f_{L2}P_{L} + f_{R2}P_{R})/M_{W}],
\vspace{0.5cm}
\end{eqnarray}
where $p_{t}$ is the four momentum of the incoming $t$ quark and $p_{b}$ is the
four momentum of the outgoing b-quark. The couplings $(f_{L1},f_{R1},f_{L2},f_{R2})$ are the vertex form
factors. In order to keep the $CP$ conservation, which is assumed in the rest of the work, 
we have the following relations among the form factors:
\begin{eqnarray}\label{formfactor}
f_{L1}^{*} = f_{L1}~ ,~~~ f_{R1}^{*} = f_{R1}~ ,~~~ f_{R2}^{*} = f_{R2}~ ,~~~ f_{L2}^{*} = f_{L2},
\end{eqnarray}
In the SM case, the coupling $f_{L1}$ is equal to one and the others are equal to zero.
The natural values of couplings $f_{L2},f_{R2}$ are in the order of $\sqrt{m_{b}m_{t}}/v\sim0.1$ \cite{Peccei}.

\section{Polarized top decay}
In the frame of SM the dominant decay chain of the top quark is: 
\begin{eqnarray}\label{topdecay}
t \rightarrow W^{+} + b \rightarrow \begin{cases}
                                     l^{+}\nu_{l} \\
                                     u\bar{d}
                                    \end{cases}
 + b
\end{eqnarray}
Because of the V-A structure of the weak interaction the decay 
products of the polarized top have a special angular correlation.
In the top rest frame, the decay angular distribution of the top products
can be written as:
\begin{eqnarray}\label{angularDis}
\frac{1}{\Gamma}\frac{d\Gamma}{d(\cos\theta_{i}^{t})} = \frac{1}{2}(1+\alpha_{i}\cos\theta_{i}^{t}),
\end{eqnarray}
where $\theta_{i}^{t}$ is the 
angle between the spin quantization axis of the top and the direction of the momentum of
the $i$th top products and $\alpha_{i}$ is called correlation coefficient or asymmetry. This correlation coefficient
shows the degree of the correlation of each decay products to the spin of the parent top quark.
In the frame of SM, it has been shown that \cite{Mahlon}:
\begin{eqnarray}\label{SM_corr}
\alpha_{l} &=& 1.0~,~\alpha_{\nu_{l}} \simeq -0.324, \nonumber \\
\alpha_{b} &=& -\frac{r^{2} - 2}{r^{2} + 2}\simeq -0.408,\nonumber \\
\alpha_{u} &\simeq& -0.324~,~\alpha_{\bar{d}} = 1.0,
\end{eqnarray}
where $r = \frac{m_{t}}{m_{W}}$. It is clear that the charged lepton is maximally correlated with its parent spin.

\subsection{Two-body decay of top quark $t\rightarrow Wb$}

The tree level  $t\rightarrow W + b$  with anomalous vertex can be simply obtained by 
using the helicity amplitude method which is available in \cite{Kane}:
\begin{eqnarray}\label{width}
\lefteqn{\Gamma_{t\rightarrow Wb} = \frac{G_{f}m_{t}m^{2}_{W}}{8\sqrt{2}\pi}
\frac{(r^{2} - 1)^{2}}{r^{4}}[(r^{2} + 2)(f_{L1}^{2}+f_{R1}^{2})}
\nonumber \\ && 
\hspace{0.5cm}+ (2r^{2} + 1)(f_{L2}^{2}+f_{R2}^{2}) + 6r(f_{L1}f_{R2} + f_{R1}f_{L2})],
\end{eqnarray}

\begin{figure}[h]
\begin{center}
\epsfig{file=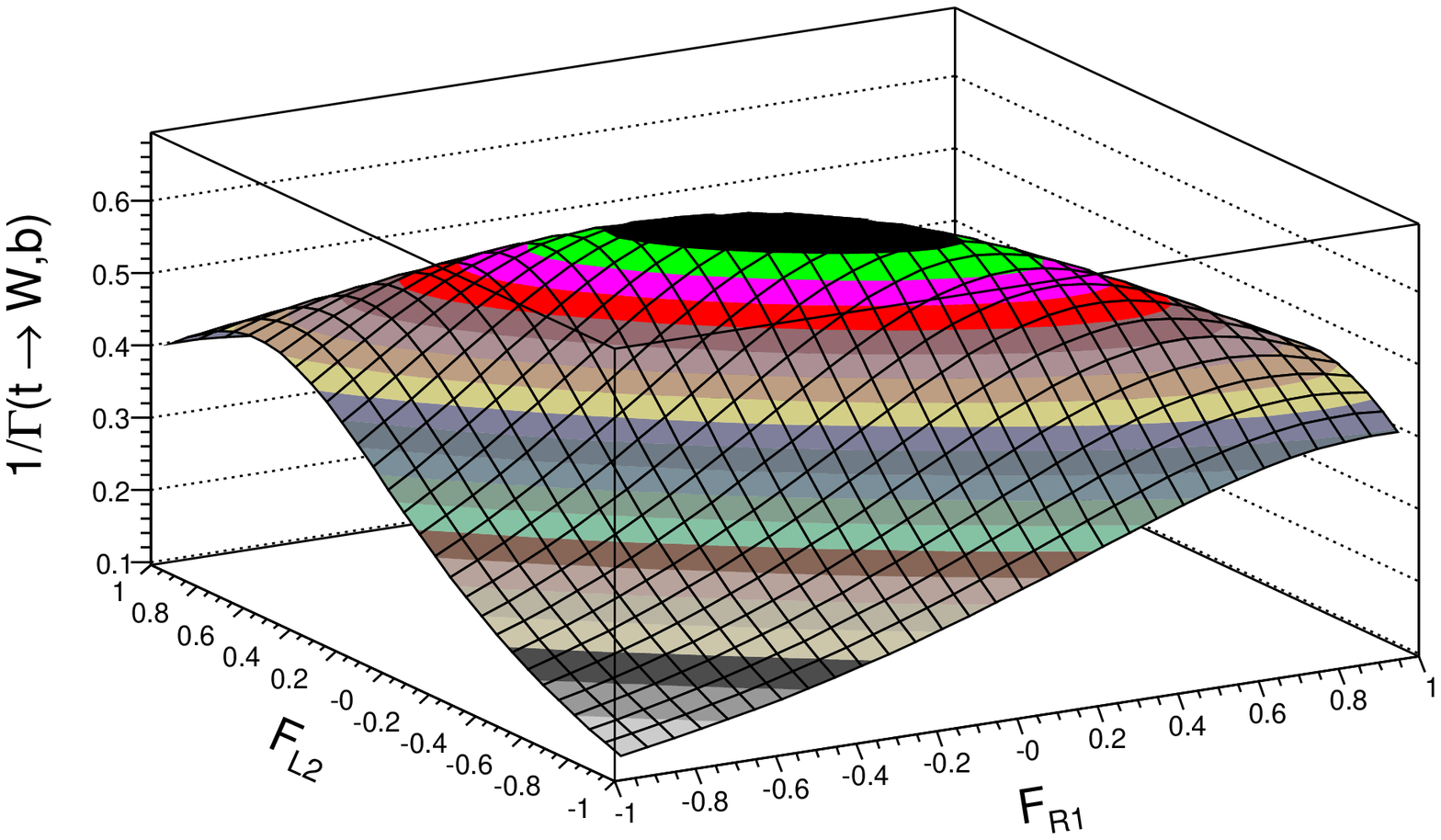,height=6.25cm,width=6.25cm}
\epsfig{file=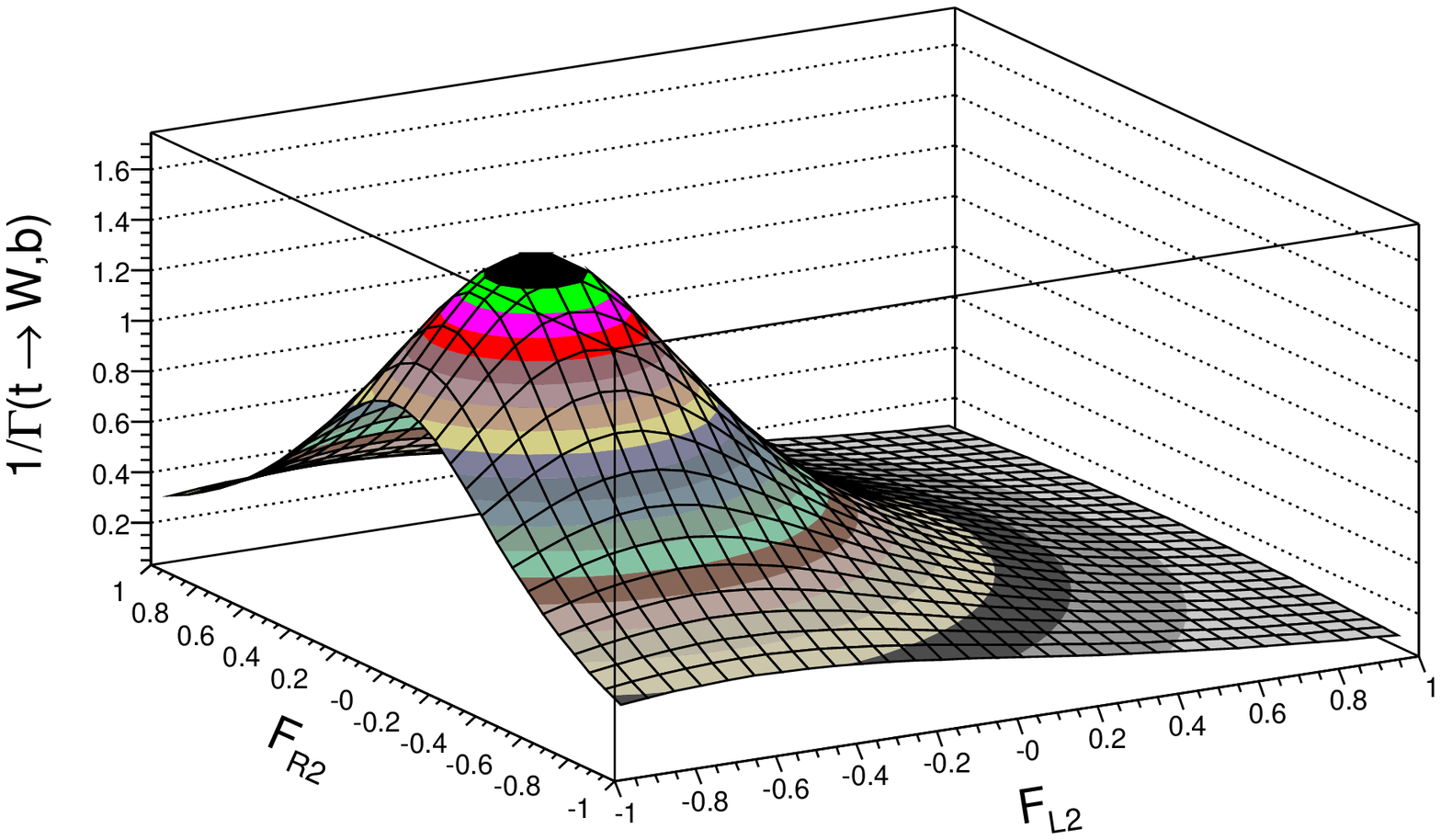,height=6.25cm,width=6.25cm}
\epsfig{file=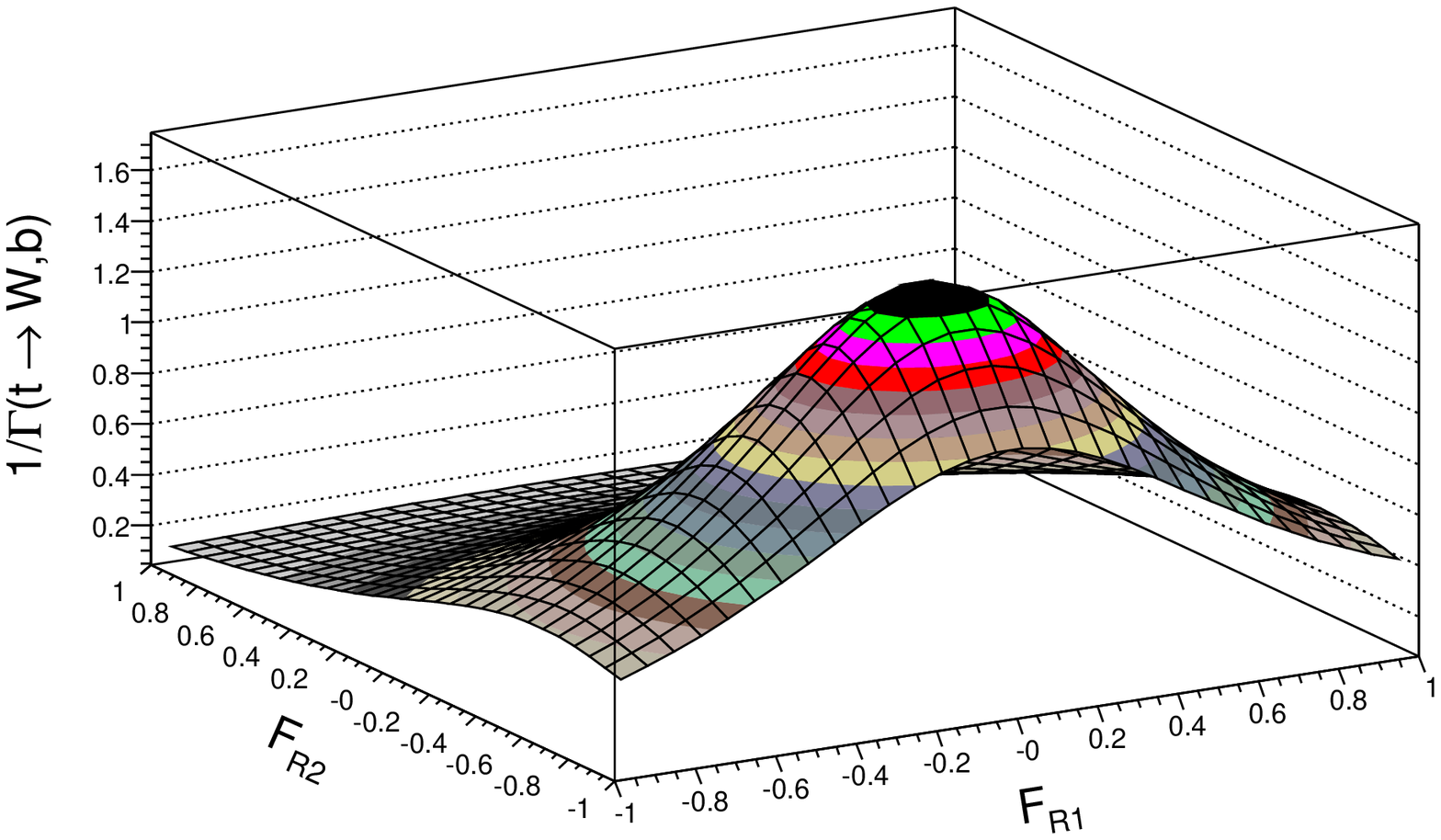,height=6.25cm,width=6.25cm}
\put(-90,0){\small{(c)}}
\put(0,200){\small{(b)}}
\put(-180,200){\small{(a)}}
 \caption{$\tau_{t}\propto 1/\Gamma_{t\rightarrow Wb}$ in terms of different anomalous couplings.} 
    \label{fig:Gamma_fr1fl2}
\end{center}
\end{figure}

It is clear that the new effective Lagrangian can change 
the width of the top quark significantly if the anomalous couplings are sufficiently large.
The variation of $\tau_{t}\propto 1/\Gamma_{t\rightarrow Wb}$ in terms of anomalous couplings are shown in two 
dimensional plots in Fig.~\ref{fig:Gamma_fr1fl2}. The Fig.(~\ref{fig:Gamma_fr1fl2}b) and Fig.(~\ref{fig:Gamma_fr1fl2}c) 
show that a small deviation in $f_{R1}$ and $f_{R2}$ leads to a non-negligible variation in the width of top quark.

The spin correlation coefficient in Eq.(\ref{angularDis}) for the b-quark 
could be calculated using the helicity amplitude approach . It has the following form in terms of anomalous couplings:
\begin{eqnarray}\label{B_coef}
\alpha_{b}&=& -\frac{(r^{2}-2)(f_{L1}^{2} - f_{R1}^{2}) + (2r^{2}-1)(f_{L2}^{2}-f_{R2}^{2})+2r(f_{R1}f_{L2}-f_{L1}f_{R2})}
{(r^{2} + 2)(f_{L1}^{2}+f_{R1}^{2})+(2r^{2} + 1)(f_{L2}^{2}+f_{R2}^{2}) + 6r(f_{L1}f_{R2} + f_{R1}f_{L2})},\nonumber \\
\end{eqnarray}
For example, $\alpha_{b}(f_{L1} = 1 , f_{R1} = f_{L2} = f_{R2} = 0.1) = -0.276$ and 
$\alpha_{b}(f_{L1} = 1 , f_{R1} = f_{L2} = f_{R2} =-0.1) = -0.548$. Comparing with
the SM case in which $\alpha_{b} = -0.408$ (see Eq.(\ref{SM_corr})), almost a big deviation is observed.
This sensitivity could be utilized in the probe of $tWb$ vertex in the experimental analyzes for obtaining
bounds on the anomalous couplings in the $t\bar{t}$ and single top events.

Applying the upper bounds (mentioned in the Introduction
for the LHC case from \cite{Boos1}) on $f_{L2}$ and $f_{R2}$ 
in $\alpha_{b}$ shows that the effect of quadratic terms 
could be important with $30\%$ which is not negligible. In particular, one should note that the $tWb$ vertex 
will be examined with high precision at the LHC in which 8 millions 
$t\bar{t}$ events will be producing per year \cite{Beneke}.

\subsection{Three-body decay of top quark $t\rightarrow l^{+}\nu_{l} b$}

The effective $tWb$ vertex in Eq.(\ref{Gamma}) contains some terms proportional to
 $\gamma_{\mu},\sigma_{\mu\nu}(p_{t}-p_{b})^{\nu}
,p_{t}^{\mu}$ and $p_{b}^{\mu}$ .
In order to simplify the calculation, we can use the Gordon identity:
\begin{eqnarray}\label{gordon}
-i\bar{b}(\sigma_{\alpha\beta}(p_{t}-p_{b})^{\beta}P_{R,L})t = 
\bar{b}(m_{t}\gamma_{\alpha}P_{L,R} - (p_{t}+p_{b})_{\alpha}P_{R,L})t,
\end{eqnarray}

In this calculation it is assumed that $m_{b}\sim 0,m_{l}\sim 0$.
Using the following definitions:
\begin{eqnarray}\label{def1}
&&~x_{l}\equiv \frac{2E_{l}}{m_{t}} = \frac{2p_{t}.p_{l}}{m_{t}^{2}}~~,~~
x_{l}\cos\theta_{l} = -\frac{2p_{l}.s}{m_{t}}~~,~~\gamma_{w}\equiv \frac{\Gamma_{W}}{m_{W}}~~,~~\nonumber\\ 
&& S_{1} \equiv \tan^{-1}[\frac{\gamma_{w}r^{2}x_{l}}{(1-r^{2}x_{l}+\gamma_{w}^{2})}]~,~
 S_{2} \equiv \log[\frac{1+\gamma_{w}^{2}}{(1-r^2x_{l})^{2}+\gamma_{w}^{2}}],
\end{eqnarray}
where $s$ is the spin four-vector of the top quark, $p_{l}$ is the four-momentum of the charged lepton, $\theta_{l}$ is 
the angle between top quark spin and the momentum of the charged lepton in the top rest frame and $\Gamma_{W}$ 
is the full width of W-boson. After following the same 
method which has been used in \cite{Kuhn}
in obtaining the lepton spectra in the polarized top quark decay the following expression is derived:
\begin{eqnarray}\label{D1}
\frac{d^{2}\Gamma_{t\rightarrow l\nu b}}{dx_{l}d\cos\theta_{l}} &=& C_{0}(T_{0}+x_{l}T_{1}+x_{l}^{2}T_{2}+x_{l}^{3}T_{3}),
\nonumber \\ 
C_{0} &=& -\frac{G_{f}^{2}m_{W}^{5}}{64\pi^{3}rx_{l}},~T_{0} = I_{R1}^{(0)}f_{R1}^2 + I_{R2}^{(0)}f_{R2}^2, \nonumber \\ 
T_{1} &=& I_{R1}^{(1)}f_{R1}^2 + I_{R2}^{(1)}f_{R2}^2 + J_{1}^{(1)}f_{R1}f_{L2}+J_{2}^{(1)}f_{R2}f_{L1},\nonumber \\ 
T_{2} &=& I_{R1}^{(2)}f_{R1}^2 + I_{R2}^{(2)}f_{R2}^2 +I_{L1}^{(2)}f_{L1}^2 + I_{L2}^{(2)}f_{L2}^2 + 
J_{1}^{(2)}f_{R1}f_{L2}+J_{2}^{(2)}f_{R2}f_{L1},\nonumber \\ 
T_{3} &=& I_{R1}^{(3)}f_{R1}^2 + I_{R2}^{(3)}f_{R2}^2 +I_{L1}^{(3)}f_{L1}^2 + I_{L2}^{(3)}f_{L2}^2,
\end{eqnarray}

where,
\begin{eqnarray}\label{D2}
I_{R1}^{(0)} &=& 4\cos\theta_{l}(S_{2}-\gamma_{w}S_{1}+S_{1}/\gamma_{w}),\nonumber \\ 
I_{R2}^{(0)} &=& 4\cos\theta_{l}(S_{2}+\gamma_{w}S_{1}-S_{1}/\gamma_{w})r^2,\nonumber \\
I_{R1}^{(1)} &=& -2\gamma_{w}S_{1}-(2+r^2)S_{2}+2S_{1}(1+r^2)/\gamma_{w}+\cos\theta_{l}(4r^2+S_{2}(2+3r^2)\nonumber\\
&&+2\gamma_{w}S_{1}-2S_{1}(1+3r^2)/\gamma_{w}),\nonumber\\
I_{R2}^{(1)} &=& 2\gamma_{w}S_{1}r^{2}+(2+r^2)r^{2}S_{2}-2(1+r^2)r^{2}S_{1}/\gamma_{w}+\cos\theta_{l}(-4r^4-S_{2}r^{2}(r^2+6)
\nonumber\\
&&-6r^{2}\gamma_{w}S_{1}+2r^{2}S_{1}(3+r^2)/\gamma_{w}),\nonumber\\
J_{1}^{(1)} &=& (-4r\gamma_{w}S_{1}-4rS_{2}+4rS_{1}/\gamma_{w})(1 + \cos\theta_{l}),\nonumber\\
J_{2}^{(1)} &=& (2r^3S_{2}-4r^3S_{1}/\gamma_{w})(1 + \cos\theta_{l} ), \nonumber\\
I_{R1}^{(2)} &=& (2(S_{2}+1)r^{2}-2S_{1}r^{2}(2+r^2)/\gamma_{w})(1 - \cos\theta_{l}),\nonumber\\
I_{R2}^{(2)} &=& -(S_{2}+2)r^{4}-r^2(2+r^2)S_{2}-2r^{2}\gamma_{w}S_{1}+2S_{1}r^{2}(1+2r^2)/\gamma_{w}\nonumber\\ 
&&+\cos\theta_{l}(6r^{4}+2r^{2}S_{2}(1+r^{2})+2r^{2}\gamma_{w}S_{1}-2r^{2}S_{1}(1+2r^2)/\gamma_{w}), \nonumber\\ 
I_{L1}^{(2)} &=& -2r^{4}S_{1}(1 + \cos\theta_{l})/\gamma_{w},\nonumber\\ 
I_{L2}^{(2)} &=& (-2r^{2}S_{2}-2r^{2}\gamma_{w}S_{1}+2r^{2}S_{1}/\gamma_{w})(1 + \cos\theta_{l}),\nonumber\\ 
J_{1}^{(2)} &=& (2r^{3}(2+S_{2}) - 4r^{3}S_{1}/\gamma_{w})(1 + \cos\theta_{l}),\nonumber\\ 
J_{2}^{(2)} &=& (-2r^{3}S_{2} + 4r^{3}S_{1}/\gamma_{w})(1 + \cos\theta_{l}),\nonumber\\ 
I_{R1}^{(3)} &=& 2r^{4}S_{1}(1 - \cos\theta_{l})/\gamma_{w},\nonumber\\ 
I_{R2}^{(3)} &=& (r^{4}(2+ S_{2}) -2r^{4}S_{1}/\gamma_{w})(1 - \cos\theta_{l}),\nonumber\\ 
I_{L1}^{(3)} &=& 2r^{4}S_{1}(1 + \cos\theta_{l})/\gamma_{w},\nonumber\\ 
I_{L2}^{(3)} &=& (r^{4}(2 + S_{2})-2r^{4}S_{1}/\gamma_{w})(1 + \cos\theta_{l}),
\end{eqnarray}
If one sets $f_{L1} = V_{tb}\simeq 1 , f_{R1} = f_{R2} = f_{L2} = 0$  in Eq.(\ref{D1}) and Eq.(\ref{D2}), the SM result 
 is obtained:
\begin{eqnarray}\label{SM}
\frac{d^{2}\Gamma_{t\rightarrow l\nu b}}{dx_{l}d\cos\theta_{l}}\bigg\vert_{SM} = 
\frac{G_{f}^{2}m_{W}^{5}r^{3}x_{l}(1-x_{l})S_{1}}
{32\pi^{3}\gamma_{w}}(1+\cos\theta_{l}),
\end{eqnarray}
which coincides with the expression which has been obtained in \cite{Kuhn}.

\begin{figure}[h]
\begin{center}
\epsfig{file=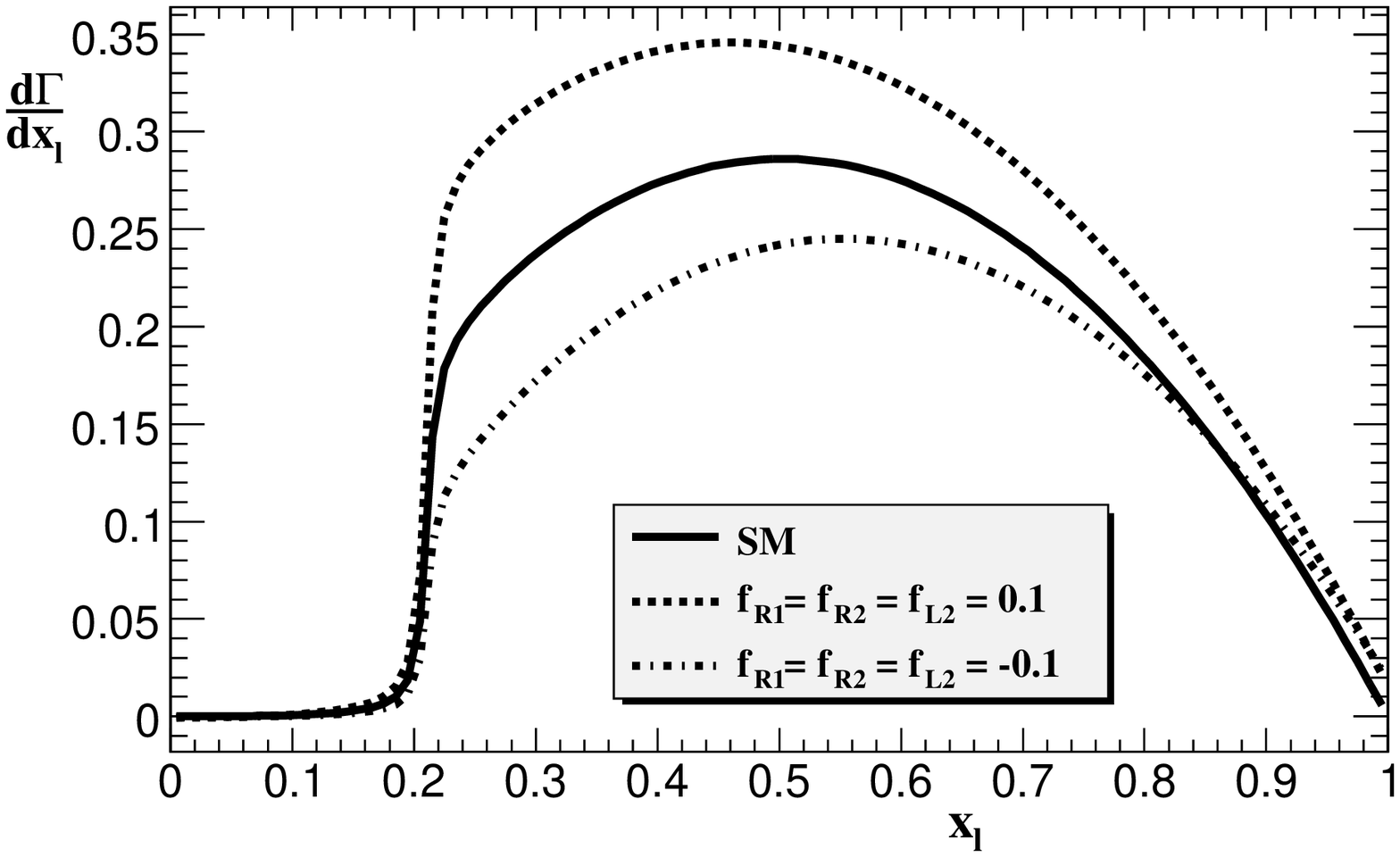,height=8cm,width=10cm}
 \caption{Energy dependence of the top quark width in terms of different anomalous couplings. Solid curve shows the SM case 
and dashed curve is for the case of $f_{L1} = 1,f_{R1} = f_{R2} = f_{L2} = 0.1$  and dash-dotted curve is for
 $f_{L1} = 1,f_{R1} = f_{R2} = f_{L2} = -0.1$.}
    \label{fig:Emu}
\end{center}
\end{figure}

\begin{figure}[h]
\begin{center}
\epsfig{file=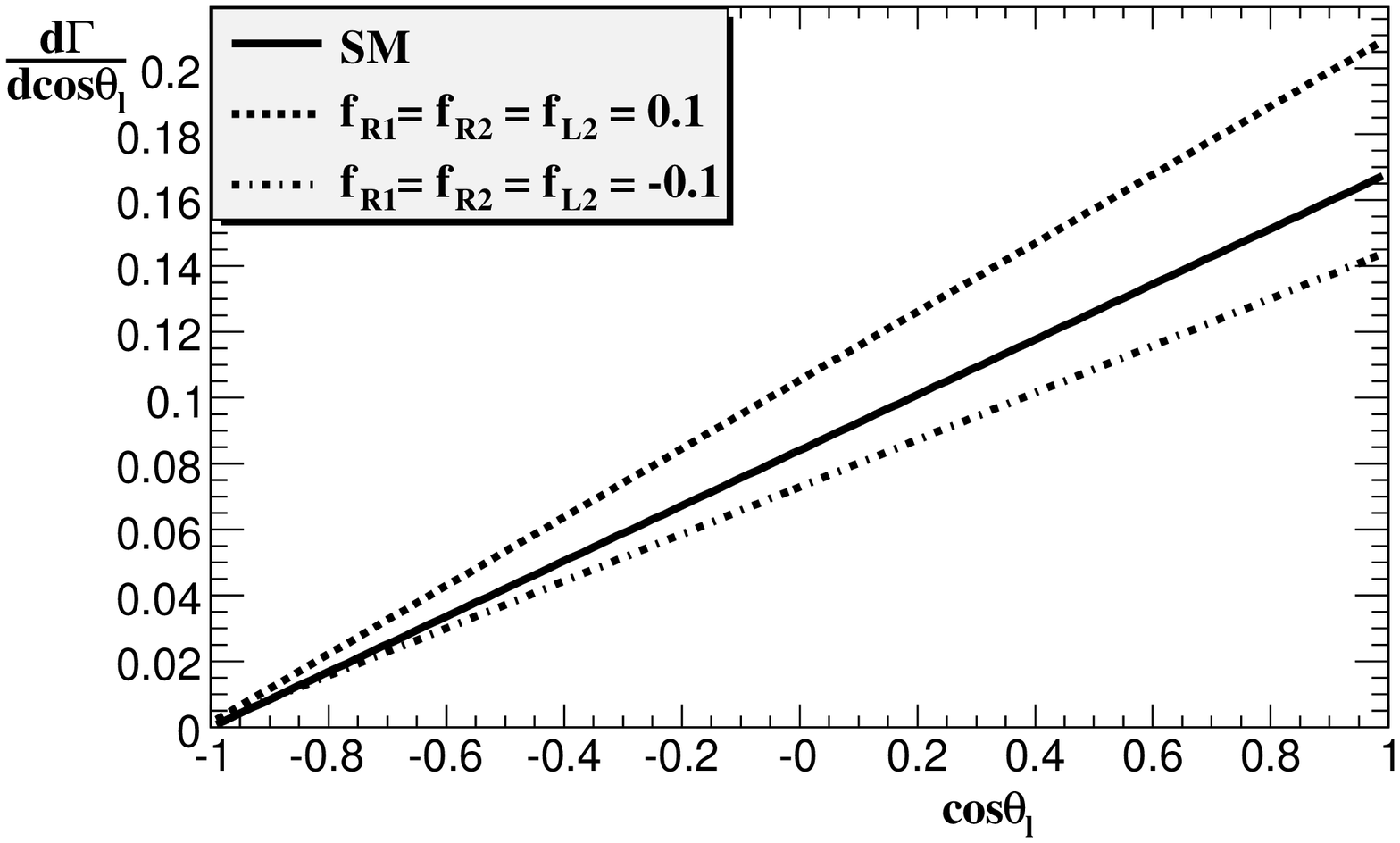,height=8cm,width=10cm}
 \caption{Angular dependence of the top quark width in terms of different anomalous couplings. Solid curve shows the SM case 
and dashed curve is for the case of $f_{L1} = 1,f_{R1} = f_{R2} = f_{L2} = 0.1$  and dash-dotted curve is for
 $f_{L1} = 1,f_{R1} = f_{R2} = f_{L2} = -0.1$.} 
    \label{fig:spin}
\end{center}
\end{figure}
In order to have a feeling of the effect of the anomalous couplings, the energy and angular dependence 
of the top quark width
are plotted by substitution of the following values in Eq.(\ref{D1}) and Eq.(\ref{D2}):
\begin{eqnarray}\label{values}
&& G_{f} = 1.166\times10^{-5}~ \text{GeV}^{-2}~~,~~m_{W} = 80.2~ \text{GeV}/\text{c}^{2}~~,\nonumber\\ 
&& m_{t} = 175~ \text{GeV}/\text{c}^{2}~~,~~\Gamma_{W} = 2.06~ \text{GeV},
\end{eqnarray}
In Fig.\ref{fig:Emu} the energy dependence of top width in SM case (solid curve) 
and $f_{L1} = 1,f_{R1} = f_{R2} = f_{L2} = 0.1$
(dashed curve)
and $f_{L1} = 1,f_{R1} = f_{R2} = f_{L2} = -0.1$
(dash-dotted curve) are plotted. The positive values for anomalous couplings lead to increment from the SM and vice versa for 
using negative values. But these deviations are mostly appeared in the range of $0.2<x_{l}<0.8$. 
Fig.\ref{fig:spin} shows three different lines with distinct slopes corresponding to SM case (solid curve) 
and $f_{L1} = 1,f_{R1} = f_{R2} = f_{L2} = 0.1$ (dashed curve) and $f_{L1} = 1,f_{R1} = f_{R2} = f_{L2} = -0.1$
(dash-dotted curve). The correlation coefficient $\alpha_{l}$ has been represented in Tab.\ref{tab:slopes}. It is clear that 
$\alpha_{l}$ receives a negligible deviation from the SM case.
\begin{table}[h]
   \begin{center}
\caption{Correlation coefficients for different values of the anomalous couplings.}
 \label{tab:slopes}
\begin{tabular}{|l|l|l|l|}
\hline
       combination  & $\Gamma(t\rightarrow l\nu_{l}b)$  & $\alpha_{l}$   &    $\alpha_{b}$ \\ \hline
 $f_{L1} = 1.0,f_{R1} = f_{L2} = f_{R2} =  0.0$ & 0.168 &    1.0         &    -0.408      \\ \hline
 $f_{L1} = 1.0,f_{R1} = f_{L2} = f_{R2} =  0.1$ & 0.210 &    0.986       &    -0.276      \\ \hline
 $f_{L1} = 1.0,f_{R1} = f_{L2} = f_{R2} = -0.1$ & 0.146 &    0.981       &    -0.548     \\ \hline
\end{tabular}
\end{center}
\end{table}

It is well known that top quarks produced in hadron colliders are scarcely polarized. As a result,
the correlations between top spin and anti-top spin in the $t\bar{t}$ is considered \cite{Beneke},\cite{Brandenburg}.
In order to illustrate we consider the decay of the $t\bar{t}$ events in hadron colliders:
\begin{eqnarray}
\bf {PP,P\bar{P}} \rightarrow t\bar{t}  \rightarrow \text{6-body final state}
\end{eqnarray}
The double differential angular distribution of the $\it{i}$th and $\it{\bar{i}}$th product coming from the top 
and anti-top in the SM is \cite{Beneke},\cite{Brandenburg}:
\begin{eqnarray}\label{spintt}
\frac{1}{\sigma}\frac{d^{2}\sigma}{d\cos\theta_{i}d\cos\bar{\theta}_{\bar{i}}} = \frac{1}{4}(1+
\kappa_{i\bar{i}} \cos\theta_{i}\cos\bar{\theta}_{\bar{i}})~,~\kappa_{i\bar{i}} = \alpha_{i}\alpha_{\bar{i}}
\times\frac{N_{\|}-N_{\times}}{N_{\|}+N_{\times}}.
\end{eqnarray}
where $\theta_{i}(\bar{\theta}_{\bar{i}})$ is the angle between the direction of the $\it{i}$th($\it{\bar{i}}$th)
product
in the rest frame of the $t(\bar{t})$ and the direction of the momentum of the $t(\bar{t})$ 
in the $t\bar{t}$ center of mass.
$N_{\|}$ is the number of top
pair events where both quarks have spin up or spin down
and $N_{\times}$ is the number of top pair events where one quark is spin up and the other is spin down.
Eq.(\ref{spintt}) shows the strong dependence of the experimental observable, $\kappa_{i\bar{i}}$, 
to the spin correlation coefficient ($\alpha_{i}$).
For the $t\bar{t}$ production at the Tevatron and the LHC,
the SM predicts $(N_{\|}-N_{\times})/(N_{\|}+N_{\times}) = 0.88~,0.33$, respectively. 
Since the $\alpha_{b}$ is very sensitive to anomalous couplings, for instance the measurement
of $\kappa_{i\bar{i}}$ with $i = b,\bar{i} = l$ might give some valuable information
concerning the non-SM couplings in hadron colliders.

\section{Conclusion}

An effective Lagrangian approach to new physics in the top quark sector was used and 
the effect of the anomalous couplings on the energy and angular dependence of a secondary charged lepton produced 
by a polarized top quark was investigated by obtaining the analytic expression.
We found that a small variation in the anomalous couplings gives a noticeable deviation in the width of top quark.
The correlation coefficient of a secondary charged lepton, which is maximally correlated with the top quark spin,
receives a negligible contribution from anomalous {\it tWb} couplings. However the correlation coefficient of the 
b-quark receives a remarkable deviation from SM. Within the current estimated bounds using  
the simulation techniques the effect of quadratic
terms of the anomalous couplings is not negligible.
The b-quark spin correlation coefficient could be used in the experimental analyzes 
in $t\bar{t}$ events to obtain better constraints on 
the anomalous couplings, specially at the LHC in which we are not restricted by statistics. 

\section{Acknowledgment}
The author would like to thank E. Boos for proposing to study this topic. Many thanks to S. Slabospitsky for his profitable
comments. Thanks to S. Paktinat Mehdiabadi for useful suggestions.

\end{document}